\documentclass[preprintnumbers,superscriptaddress,nofootinbib,aps,prd,twocolumn,floatfix]{revtex4}
\pdfoutput=1

\usepackage{multirow,array}
\usepackage{amsmath,amssymb}
\usepackage{graphicx}
\usepackage{color}
\usepackage{url}
\usepackage{feynmp}
\usepackage{slashed}
\usepackage{multirow}
\usepackage{footmisc}
\usepackage{hyperref}
\usepackage[scientific-notation=true]{siunitx}
\usepackage{dsfont}

\usepackage{pgf,tikz}
\usetikzlibrary{
  patterns,
  shapes.multipart,
  arrows,
  trees,
  scopes,
  decorations.pathreplacing,
  decorations.pathmorphing,
  decorations.markings,
  decorations.text,
  positioning,
  calc
}
\tikzset{
  dirac/.style=
  {
    draw=black,
    postaction={decorate},
    decoration=
    {
      markings,
      mark=at position .6 with {\arrow[thick,draw=black]{>}}
    }
  },
  vector/.style=
  {
    decorate,
    draw=black,
    decoration={snake,amplitude=1.5pt,segment length=3pt}
  },
  real_scalar/.style=
  {
    densely dashed,
    thick,
    draw=black,
  },
  scalar/.style=
  {
    densely dashed,
    thick,
    draw=black,
  },
  gluon/.style=
  {
    decorate,
    draw=black,
    decoration={coil,amplitude=2pt,segment length=2.5pt}
  },
  vertex/.style=
  {
    on grid,
    draw=black,
    fill=black,
    circle,
    minimum size=2pt, 
    inner sep=0pt
  },
  eff_vertex/.style=
  {
    on grid,
    draw=black,
    circle,
    fill=black,
    fill opacity=0.2,
    minimum size=5pt,
    inner sep=1pt
  }
}

\hypersetup{
  pdfauthor={Dorival Goncalves, Frank Krauss, Silvan Kuttimalai, Philipp Maierhoefer},
  pdftitle={Boosting invisible searches via $ZH$: From the Higgs Boson to Dark Matter Simplified Models},
  pdfkeywords={QCD, NLO, Higgs Physics}
}
\begin{document}
\def\contentsname{{\normalsize Content}}
\def\tablename{Table}
\def\figurename{Figure}

\def\bdrs{\text{BDRS}}
\def\pveto{P_\text{veto}}
\def\nj{n_\text{jets}}
\def\meff{m_\text{eff}}
\def\ptmin{p_T^\text{min}}
\def\gtot{\Gamma_\text{tot}}
\def\as{\alpha_s}
\def\az{\alpha_0}
\def\gz{g_0}
\def\w{\vec{w}}
\def\sdag{\Sigma^{\dag}}
\def\s{\Sigma}
\newcommand{\psib}{\overline{\psi}}
\newcommand{\Psib}{\overline{\Psi}}
\newcommand\one{\leavevmode\hbox{\small1\normalsize\kern-.33em1}}
\newcommand{\Mpl}{M_\mathrm{Pl}}
\newcommand{\p}{\partial}
\newcommand{\mat}{\mathcal{M}}
\newcommand{\lag}{\mathcal{L}}
\newcommand{\ord}{\mathcal{O}}
\newcommand{\ope}{\mathcal{O}}
\newcommand{\qqquad}{\qquad \qquad}
\newcommand{\qqqquad}{\qquad \qquad \qquad}

\newcommand{\qb}{\bar{q}}
\newcommand{\matx}{|\mathcal{M}|^2}
\newcommand{\really}{\stackrel{!}{=}}
\newcommand{\msbar}{\overline{\text{MS}}}
\newcommand{\qns}{f_q^\text{NS}}
\newcommand{\lqcd}{\Lambda_\text{QCD}}
\newcommand{\met}{E_T^{\text{miss}}}
\newcommand{\pmiss}{\slashchar{\vec{p}}_T}

\newcommand{\sq}{\tilde{q}}
\newcommand{\go}{\tilde{g}}
\newcommand{\st}[1]{\tilde{t}_{#1}}
\newcommand{\stb}[1]{\tilde{t}_{#1}^*}
\newcommand{\nz}[1]{\tilde{\chi}_{#1}^0}
\newcommand{\cp}[1]{\tilde{\chi}_{#1}^+}
\newcommand{\CP}{CP}

\providecommand{\mg}{m_{\tilde{g}}}
\providecommand{\mst}[1]{m_{\tilde{t}_{#1}}}
\newcommand{\msn}[1]{m_{\tilde{\nu}_{#1}}}
\newcommand{\mch}[1]{m_{\tilde{\chi}^+_{#1}}}
\newcommand{\mne}[1]{m_{\tilde{\chi}^0_{#1}}}
\newcommand{\msb}[1]{m_{\tilde{b}_{#1}}}
\newcommand{\vsm}{\ensuremath{v_{\rm SM}}}

\newcommand{\mev}{{\ensuremath\rm MeV}}
\newcommand{\gev}{{\ensuremath\rm GeV}}
\newcommand{\tev}{{\ensuremath\rm TeV}}
\newcommand{\sign}{{\ensuremath\rm sign}}
\newcommand{\iab}{{\ensuremath\rm ab^{-1}}}
\newcommand{\ifb}{{\ensuremath\rm fb^{-1}}}
\newcommand{\ipb}{{\ensuremath\rm pb^{-1}}}

\def\slashchar#1{\setbox0=\hbox{$#1$}           
   \dimen0=\wd0                                 
   \setbox1=\hbox{/} \dimen1=\wd1               
   \ifdim\dimen0>\dimen1                        
      \rlap{\hbox to \dimen0{\hfil/\hfil}}      
      #1                                        
   \else                                        
      \rlap{\hbox to \dimen1{\hfil$#1$\hfil}}   
      /                                         
   \fi}
\newcommand{\dslash}{\slashchar{\partial}}
\newcommand{\Dslash}{\slashchar{D}}

\newcommand{\eg}{\textsl{e.g.}\;}
\newcommand{\ie}{\textsl{i.e.}\;}
\newcommand{\etal}{\textsl{et al}\;}

\setlength{\floatsep}{0pt}
\setcounter{topnumber}{1}
\setcounter{bottomnumber}{1}
\setcounter{totalnumber}{1}
\renewcommand{\topfraction}{1.0}
\renewcommand{\bottomfraction}{1.0}
\renewcommand{\textfraction}{0.0}
\renewcommand{\thefootnote}{\fnsymbol{footnote}}

\newcommand{\rig}{\rightarrow}
\newcommand{\lrig}{\longrightarrow}
\renewcommand{\d}{{\mathrm{d}}}
\newcommand{\be}{\begin{eqnarray*}}
\newcommand{\ee}{\end{eqnarray*}}
\newcommand{\gl}[1]{(\ref{#1})}
\newcommand{\ta}[2]{ \frac{ {\mathrm{d}} #1 } {{\mathrm{d}} #2}}
\newcommand{\bee}{\begin{eqnarray}}
\newcommand{\eee}{\end{eqnarray}}
\newcommand{\beeq}{\begin{equation}}
\newcommand{\eeeq}{\end{equation}}
\newcommand{\mc}{\mathcal}
\newcommand{\mr}{\mathrm}
\newcommand{\ep}{\varepsilon}
\newcommand{\emt}{$\times 10^{-3}$}
\newcommand{\emfo}{$\times 10^{-4}$}
\newcommand{\emfi}{$\times 10^{-5}$}

\newcommand{\revision}[1]{{\bf{}#1}}

\newcommand{\hzero}{h^0}
\newcommand{\Hzero}{H^0}
\newcommand{\Azero}{A^0}
\newcommand{\PHiggs}{H}
\newcommand{\PW}{W}
\newcommand{\PZ}{Z}

\newcommand{\sw}{\ensuremath{s_w}}
\newcommand{\cw}{\ensuremath{c_w}}
\newcommand{\swd}{\ensuremath{s^2_w}}
\newcommand{\cwd}{\ensuremath{c^2_w}}

\newcommand{\mhhd}{\ensuremath{m^2_{\Hzero}}}
\newcommand{\mhh}{\ensuremath{m_{\Hzero}}}
\newcommand{\mlhd}{\ensuremath{m^2_{\hzero}}}
\newcommand{\Mlh}{\ensuremath{m_{\hzero}}}
\newcommand{\mad}{\ensuremath{m^2_{\Azero}}}
\newcommand{\mhpd}{\ensuremath{m^2_{\PHiggs^{\pm}}}}
\newcommand{\mhp}{\ensuremath{m_{\PHiggs^{\pm}}}}

 \newcommand{\sa}{\ensuremath{\sin\alpha}}
 \newcommand{\ca}{\ensuremath{\cos\alpha}}
 \newcommand{\cad}{\ensuremath{\cos^2\alpha}}
 \newcommand{\sad}{\ensuremath{\sin^2\alpha}}
 \newcommand{\sbd}{\ensuremath{\sin^2\beta}}
 \newcommand{\cbd}{\ensuremath{\cos^2\beta}}
 \newcommand{\cb}{\ensuremath{\cos\beta}}
 \renewcommand{\sb}{\ensuremath{\sin\beta}}
 \newcommand{\tanbd}{\ensuremath{\tan^2\beta}}
 \newcommand{\cotbd}{\ensuremath{\cot^2\beta}}
 \newcommand{\tanb}{\ensuremath{\tan\beta}}
 \newcommand{\tb}{\ensuremath{\tan\beta}}
 \newcommand{\cotb}{\ensuremath{\cot\beta}}



\title{Boosting invisible searches via $\boldsymbol{ZH}$:
\\From the Higgs Boson to Dark Matter Simplified Models}

\author{Dorival Gon\c{c}alves}
\author{Frank Krauss}
\author{Silvan Kuttimalai}
\affiliation{
  Institute for Particle Physics Phenomenology,
  Physics Department, Durham University,
  Durham DH1 3LE, United Kingdom}
\author{Philipp Maierh\"ofer}
\affiliation{
  Physikalisches Institut,
  Albert-Ludwigs-Universit\"at Freiburg,
  79104 Freiburg,
  Germany}

\begin{abstract}
  \noindent
  Higgs boson production in association with a $Z$-boson at the LHC is
  analysed, both in the Standard Model and in Simplified Model extensions for
  Dark Matter.  We focus on $H\rightarrow$~\textit{invisibles} searches
  and show that loop-induced components for both the signal and background
  present phenomenologically relevant contributions to the
  $\mathcal{BR}(H\rightarrow\textit{inv})$ limits.  In addition, the 
  constraining power of this channel to Simplified Models for Dark Matter
  with scalar and pseudo-scalar mediators $\phi$ and $A$ is discussed and
  compared with non-collider constraints.  We find that with $100~fb^{-1}$ of
  LHC data, this channel provides competitive constraints to the non-collider
  bounds, for most of the parameter space we consider, bounding the universal
  Standard Model fermion-mediator strength at $g_v < 1$ for moderate masses
  in the range of ${100~\text{GeV}<m_{\phi/A}<400}$~GeV.
\end{abstract}

\maketitle

\tableofcontents

\section{Introduction}
\label{sec:intro}

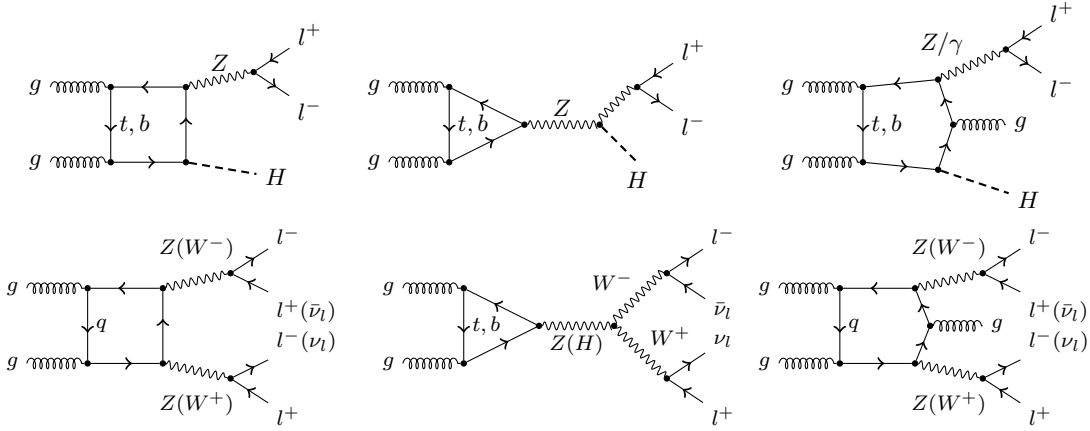
\begin{figure*}
  \centering
  {\begin{tikzpicture}[]
 \node [on grid] (gu) at (-1.5,+0.5) {$g$};
 \node [on grid] (gd) at (-1.5,-0.5) {$g$};

 \node [vertex] (tul) at (-0.5,+0.5) {};
 \node [vertex] (tdl) at (-0.5,-0.5) {};
 \node [vertex] (tur) at (+0.5,+0.5) {};
 \node [vertex] (tdr) at (+0.5,-0.5) {};
 \draw [gluon]  (gu)  -- (tul);
 \draw [gluon]  (gd)  -- (tdl);
 \draw [dirac]  (tul) -- (tdl) node [midway, right] {$t,b$};
 \draw [dirac]  (tdl) -- (tdr);
 \draw [dirac]  (tdr) -- (tur);
 \draw [dirac]  (tur) -- (tul);

 \node [vertex] (zu)  at ($(tur)+(0.9,+0.2)$) {};
 \node [      ] (zd)  at ($(tdr)+(1.2,-0.2)$) {$H$};
 \draw [vector] (tur) -- (zu) node [midway, above] {\small $Z$};
 \draw [scalar] (tdr) -- (zd);

 \node [on grid] (lu1)  at ($(zu)+(0.75,+0.5)$) {$l^+$};
 \node [on grid] (lu2)  at ($(zu)+(0.75,-0.5)$) {$l^-$};
 \draw [dirac]   (lu1) -- (zu);
 \draw [dirac]   (zu) -- (lu2);

 \begin{scope}[shift={(4.5,0)}]
   \node [on grid] (gu) at (-1.5,+0.5) {$g$};
   \node [on grid] (gd) at (-1.5,-0.5) {$g$};
   
   \node [vertex] (tu) at (-0.5,+0.5) {};
   \node [vertex] (td) at (-0.5,-0.5) {};
   \node [vertex] (t)  at (+0.5,+0.0) {};
   \draw [gluon]  (gu)  -- (tu);
   \draw [gluon]  (gd)  -- (td);
   \draw [dirac]  (tu)  -- (td) node [midway, right] {$t,b$};
   \draw [dirac]  (td)  -- (t) ;
   \draw [dirac]  (t)   -- (tu);

   \node [vertex] (zv) at ($(t)+(1.0,0.0)$) {};
   \draw [vector] (t)  -- (zv) node [midway,above] {$Z$} --++ (.5,0.5) node (z) [vertex] {};
   \node [on grid] (lu1)  at ($(z)+(0.75,+0.5)$) {$l^+$};
   \node [on grid] (lu2)  at ($(z)+(0.75,-0.5)$) {$l^-$};
   \draw [dirac]   (lu1) -- (z);
   \draw [dirac]   (z)   -- (lu2);
   
   \draw [scalar] (zv) --++ (0.5,-0.5) node (h) [below] {$H$};
 \end{scope}

 \begin{scope}[shift={(10,0)}]
   \node [on grid] (gu) at (-1.5,+0.5) {$g$};
   \node [on grid] (gd) at (-1.5,-0.5) {$g$};

   \node [vertex] (tul) at (-0.5,+0.5) {};
   \node [vertex] (tdl) at (-0.5,-0.5) {};
   \node [vertex] (tur) at (+0.5,+0.6) {};
   \node [vertex] (tr)  at (+0.7,+0.0) {};
   \node [vertex] (tdr) at (+0.5,-0.6) {};
   \draw [gluon]  (gu)  -- (tul);
   \draw [gluon]  (gd)  -- (tdl);
   \draw [dirac]  (tul) -- (tdl) node [midway, right] {$t,b$};
   \draw [dirac]  (tdl) -- (tdr);
   \draw [dirac]  (tdr) -- (tr);
   \draw [dirac]  (tr)  -- (tur);
   \draw [dirac]  (tur) -- (tul);

   \node [vertex] (zu)  at ($(tur)+(0.9,+0.4)$) {};
   \node [      ] (zd)  at ($(tdr)+(1.2,-0.4)$) {$H$};
   \draw [vector] (tur) -- (zu) node [midway, above left] {\small $Z/\gamma$};
   \draw [scalar] (tdr) -- (zd);

   \node [on grid] (lu1)  at ($(zu)+(0.75,+0.5)$) {$l^+$};
   \node [on grid] (lu2)  at ($(zu)+(0.75,-0.5)$) {$l^-$};
   \draw [dirac]   (lu1) -- (zu);
   \draw [dirac]   (zu)  -- (lu2);

   \draw [gluon] (tr) --++ (0.7,0.0) node [right] {$g$};
 \end{scope}
\end{tikzpicture}
}
  {\begin{tikzpicture}[every node/.style={font={\fontsize{8pt}{12}\selectfont}}]
 \node [on grid] (gu) at (-1.5,+0.5) {$g$};
 \node [on grid] (gd) at (-1.5,-0.5) {$g$};

 \node [vertex] (tul) at (-0.5,+0.5) {};
 \node [vertex] (tdl) at (-0.5,-0.5) {};

 \node [vertex] (tur) at (+0.5,+0.5) {};
 \node [vertex] (tdr) at (+0.5,-0.5) {};

 \node [vertex] (zu)  at ($(tur)+(0.9,+0.2)$) {};
 \node [vertex] (zd)  at ($(tdr)+(0.9,-0.2)$) {};

 \node [on grid, anchor=west] (lu1)  at ($(zu)+(0.5,+0.5)$) {$l^-$};
 \node [on grid, anchor=west] (lu2)  at ($(zu)+(0.5,-0.5)$) {$l^+(\bar\nu_l)$};

 \node [on grid, anchor=west] (nd1)  at ($(zd)+(0.5,+0.5)$) {$l^-(\nu_l)$};
 \node [on grid, anchor=west] (nd2)  at ($(zd)+(0.5,-0.5)$) {$l^+$};

 \draw [gluon]  (gu)  -- (tul);
 \draw [gluon]  (gd)  -- (tdl);
 \draw [dirac]  (tul) -- (tdl) node [midway, right] {$q$};
 \draw [dirac]  (tdl) -- (tdr);
 \draw [dirac]  (tdr) -- (tur);
 \draw [dirac]  (tur) -- (tul);

 \draw [vector] (tur) -- (zu) node [midway, above, outer sep=5] {$Z(W^-)$};
 \draw [vector] (tdr) -- (zd) node [midway, below, outer sep=5] {$Z(W^+)$};

 \draw [dirac]  (zu)  -- (lu1);
 \draw [dirac]  (lu2) -- (zu) ;

 \draw [dirac]  (zd)  -- (nd1);
 \draw [dirac]  (nd2) -- (zd) ;
  
 \begin{scope}[shift={(5,0)}]
   \node [on grid] (gu) at (-1.5,+0.5) {$g$};
   \node [on grid] (gd) at (-1.5,-0.5) {$g$};
   
   \node [vertex] (tu) at (-0.5,+0.5) {};
   \node [vertex] (td) at (-0.5,-0.5) {};
   \node [vertex] (t)  at (+0.5,+0.0) {};
   \draw [gluon]  (gu)  -- (tu);
   \draw [gluon]  (gd)  -- (td);
   \draw [dirac]  (tu)  -- (td) node [midway, right] {$t,b$};
   \draw [dirac]  (td)  -- (t) ;
   \draw [dirac]  (t)   -- (tu);

   \node [vertex] (zv) at ($(t)+(1.0,0.0)$) {};
   \draw [vector] (t)  -- (zv) node [midway,below] {$Z(H)$} --++ (.7,0.7)
   node [midway, above left] {$W^-$} node (z) [vertex] {};
   \node [on grid] (lu1)  at ($(z)+(0.75,+0.5)$) {$l^-$};
   \node [on grid] (lu2)  at ($(z)+(0.75,-0.5)$) {$\bar\nu_l$};
   \draw [dirac]  (z)   -- (lu1);
   \draw [dirac]  (lu2) -- (z);
   
   \draw [vector] (zv) --++ (0.7,-0.7) node [midway, above right] {$W^+$} node [vertex] (w) {};
   \node [on grid] (ld1)  at ($(w)+(0.75,+0.5)$) {$\nu_l$};
   \node [on grid] (ld2)  at ($(w)+(0.75,-0.5)$) {$l^+$};
   \draw [dirac]  (w)  -- (ld1);
   \draw [dirac]  (ld2) -- (w);

 \end{scope}

 \begin{scope}[shift={(10.0,0)}]
   \node [on grid] (gu) at (-1.5,+0.5) {$g$};
   \node [on grid] (gd) at (-1.5,-0.5) {$g$};

   \node [vertex] (tul) at (-0.5,+0.5) {};
   \node [vertex] (tdl) at (-0.5,-0.5) {};
   \node [vertex] (tur) at (+0.5,+0.5) {};
   \node [vertex] (tr)  at (+0.7,+0.0) {};
   \node [vertex] (tdr) at (+0.5,-0.5) {};
   \draw [gluon]  (gu)  -- (tul);
   \draw [gluon]  (gd)  -- (tdl);
   \draw [dirac]  (tul) -- (tdl) node [midway, right] {$q$};
   \draw [dirac]  (tdl) -- (tdr);
   \draw [dirac]  (tdr) -- (tr);
   \draw [dirac]  (tr)  -- (tur);
   \draw [dirac]  (tur) -- (tul);

   \node [vertex] (zu)  at ($(tur)+(0.9,+0.2)$) {};
   \node [vertex] (zd)  at ($(tdr)+(0.9,-0.2)$) {};
   \draw [vector] (tur) -- (zu) node [midway, above, outer sep=5] {$Z(W^-)$};
   \draw [vector] (tdr) -- (zd) node [midway, below, outer sep=5] {$Z(W^+)$};
   \node [on grid, anchor=west] (lu1)  at ($(zu)+(0.5,+0.5)$) {$l^-$};
   \node [on grid, anchor=west] (lu2)  at ($(zu)+(0.5,-0.5)$) {$l^+(\bar\nu_l)$};
   \node [on grid, anchor=west] (nd1)  at ($(zd)+(0.5,+0.5)$) {$l^-(\nu_l)$};
   \node [on grid, anchor=west] (nd2)  at ($(zd)+(0.5,-0.5)$) {$l^+$};
   \draw [dirac]  (zu)  -- (lu1);
   \draw [dirac]  (lu2) -- (zu) ;
   \draw [dirac]  (zd)  -- (nd1);
   \draw [dirac]  (nd2) -- (zd) ;

   \draw [gluon] (tr) --++ (0.7,0.0) node [right] {$g$};
\end{scope}

\end{tikzpicture}
}
  \caption{
    Top panel: Representative loop-induced Feynman diagrams contributing to
    the signal $Z(ll)H$ (left and central) and $Z(ll)Hj$ (right).
    Bottom panel: The same for the background $ZZ$ and $WW$
    (left and central) and $ZZj,WWj$ (right). In the massless limit for first
    and second generation quarks, only top and bottom flavours contribute
    to the triangle graphs as a consequence of Furry's theorem.}
  \label{fig:diags}
\end{figure*}

Constraining the invisible decay width of the Higgs boson is a cornerstone
of the LHC programme~\cite{higgs,discovery, inv_exp,legacy,inv_wbf}.  The
Standard Model (SM) predicts a very small Higgs to invisible decay width
which is practically inaccessible with the LHC sensitivity.  However, many
models collectively referred to as Higgs portal models predict a larger
invisible branching ratio~\cite{hportal,Bernaciak:2014pna}.  The main hope
of these models is to establish a link to a potential Dark Sector through
the Higgs boson.  Therefore, any determination of the invisible decay width
of the Higgs boson would directly represent a new physics discovery and
could be connected to a Dark Matter (DM) candidate.

One of the most prominent and phenomenologically stringent LHC invisible Higgs
search channels is Higgs-Strahlung, $ZH$ production. The associated $Z$ and
Higgs boson production generates an interesting signature, characterised by
a boosted di-leptonic pair recoiling against large missing transverse energy.
The current upper bounds derived with this channel by the LHC experiments are
${\mathcal{BR}(H\rightarrow inv)<0.75}$ and $0.58$ at 95\% CL for ATLAS and
CMS, respectively~\cite{inv_exp}.\medskip

In this contribution, the Higgs-Strahlung channel is carefully explored,
emphasising the fundamental ingredients for a robust theoretical prediction.
In particular, we show that both the loop-induced signal $ZH$ and backgrounds
$WW$ and $ZZ$ play a fundamental role. To our knowledge, this is the first
dedicated study that scrutinises the importance of the latter in the
considered search.  We also show their impact in a full signal--background
study deriving the LHC $\mathcal{BR}(H\rightarrow inv)$ limits.

The present study is further extended to a set of Simplified Models for Dark
Matter.  In these models, DM is produced through a scalar $\phi$ or a
pseudo-scalar $A$ mediator that produces relevant rates through the
loop-induced $Z\phi(A)$ channel. We derive the LHC sensitivity as a function
of the mediator mass.
\medskip

This paper is organised as follows. In Sec.~\ref{sec:zh_loop}, the importance
of the loop-induced components to signal and background is quantified, and
the importance of multi-jet merging techniques to the current Higgs-Strahlung
searches is discussed. In Sec.~\ref{sec:inv}, a complete signal--background
analysis is presented for searches for invisible Higgs decays, and we derive
the sensitivity of the current Run II at the LHC.  In Sec.~\ref{sec:DM}, we
further extend this analysis, exploring the Simplified Models for Dark Matter
via the $Z\phi(A)$ channel. The summary of our results is
presented in Sec.~\ref{sec:summary}.

\section{Ingredients of the analysis}
\label{sec:zh_loop}

\subsection{Loop-induced signal and background}

\begin{figure*}[tbp]
  \includegraphics[width=.48\textwidth]{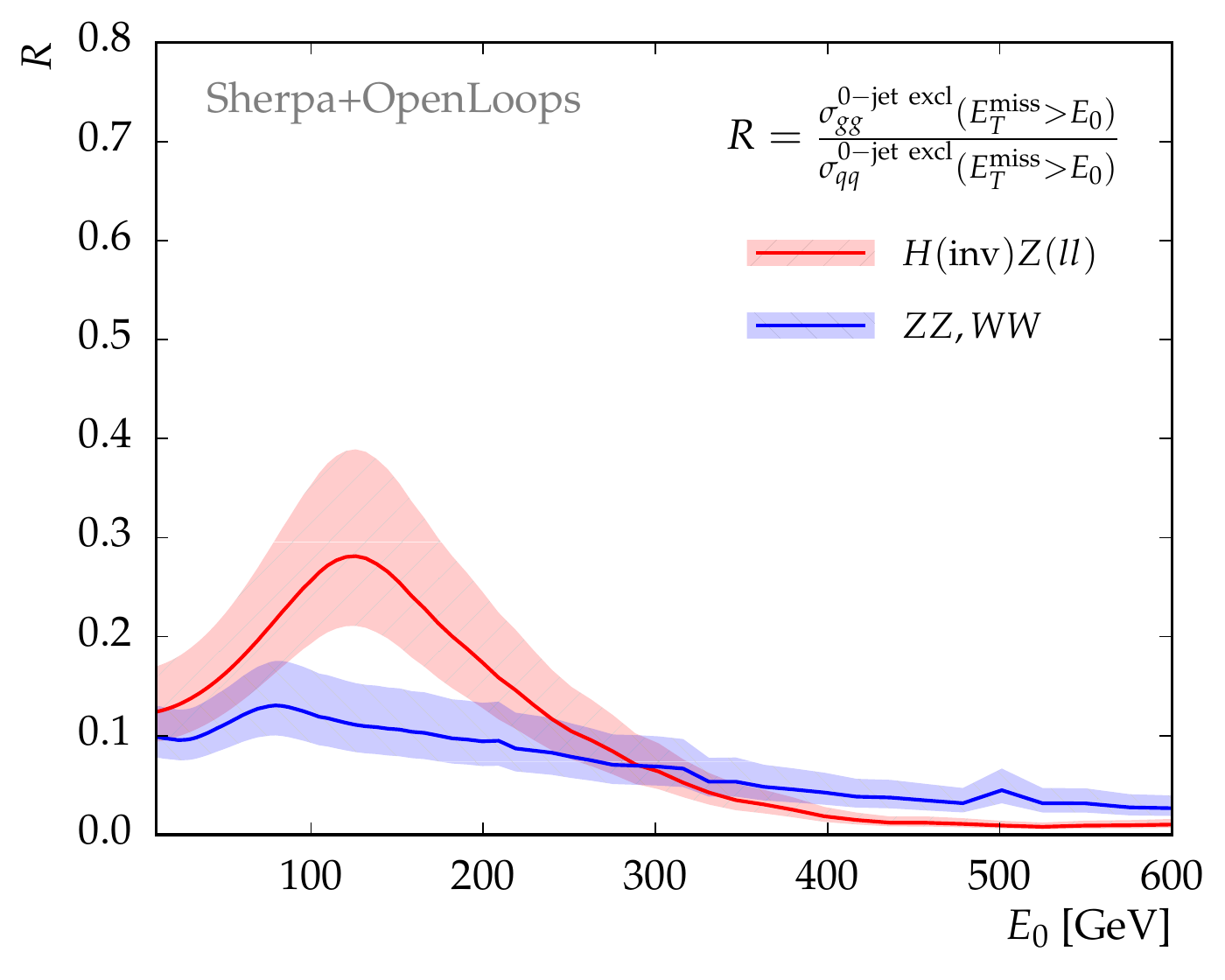}
  \hspace{0.5cm}
  \includegraphics[width=.48\textwidth]{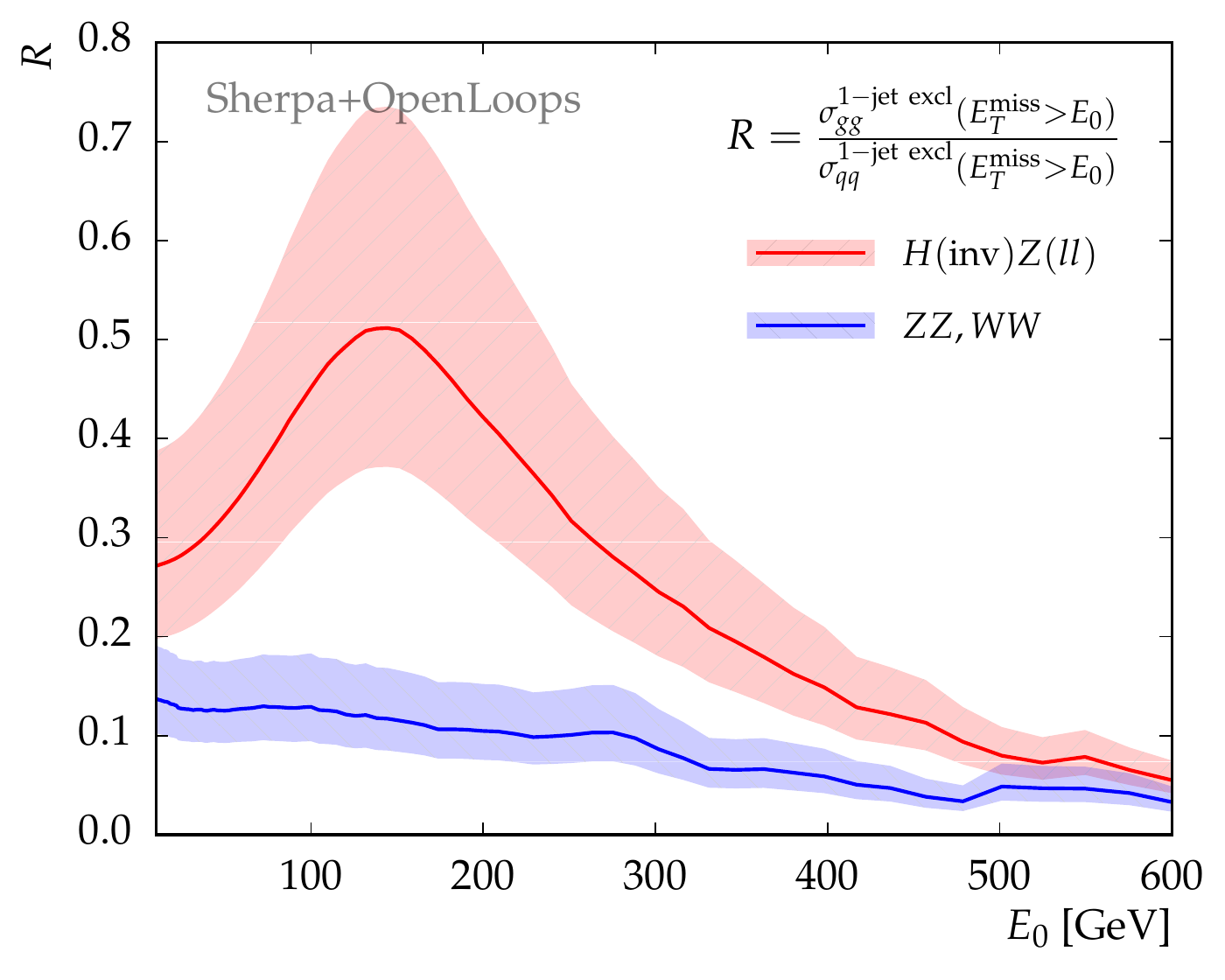}
  \caption{
    The fraction of the gluon fusion contribution to the Higgs-Strahlung
    cross-section, for the $Z(ll)H(\text{inv})$ signal as well as the
    $V^{(*)}V'^{(*)}=Z(ll)Z(\nu\nu),W(l\nu)W(l\nu)$ background, as a function
    of the $\slashed{E}_T>E_0$ selection cut in the zero-jet (left panel)
    and one-jet (right panel) exclusive bins.  Spin correlations and
    off-shell effects are fully accounted for in the vector boson decays.
    The NLO Drell-Yan and the loop-induced gluon fusion samples,
    for signal and background, are merged up to one jet, at next-to leading
    order and at leading order, respectively. The uncertainty bands result
    from 3-point scale variations on the matrix element.}
  \label{fig:met}
\end{figure*}

The Higgs-Strahlung signal $Z(ll)H(inv)$ and dominant background
$VV'=Z(ll)Z(\nu\nu),W(l\nu)W(l\nu)$ present structural similarities relevant
for the invisible searches. Both are dominated, at the level of total rates,
by the quark initiated sub-processes, which we refer to as Drell-Yan-like
(DY), and indicate them with the subscript $qq$.  In both cases, $ZH$ and
$VV'$, there are also loop-induced gluon fusion (GF) contributions, indicated
by the subscript $gg$, that become important in some kinematic regimes
despite their sub-leading corrections to the total rate~\cite{zh_ours,
  zh_gf,zh_merging_DY,zh_merging_GF,harlander1,nlo_ew,harlander,harlander2,
  WW_+_jets}.  It is clear that this classification, strictly speaking is
valid at Born-level only; higher-order corrections of course also include
different initial states.  In addition, the loop--\-induced contributions
are part of the NNLO correction to the process, which do not interfere
with the other contributions at this order.  However, current event
simulation technology is not yet able to include these contributions in a
more systematic way, and they have to be added as independent samples.\footnote{Similarly
to  the Higgs pair production, higher-order QCD effects result in  large corrections
to the considered GF processes. The GF rates account for $K=2$~\cite{zh_ours,zh_gf}.} 
In Fig.~\ref{fig:diags}, we display a representative sample of the GF Feynman
diagrams for both signal and background.

\begin{figure*}[!t]
  \includegraphics[width=.48\textwidth]{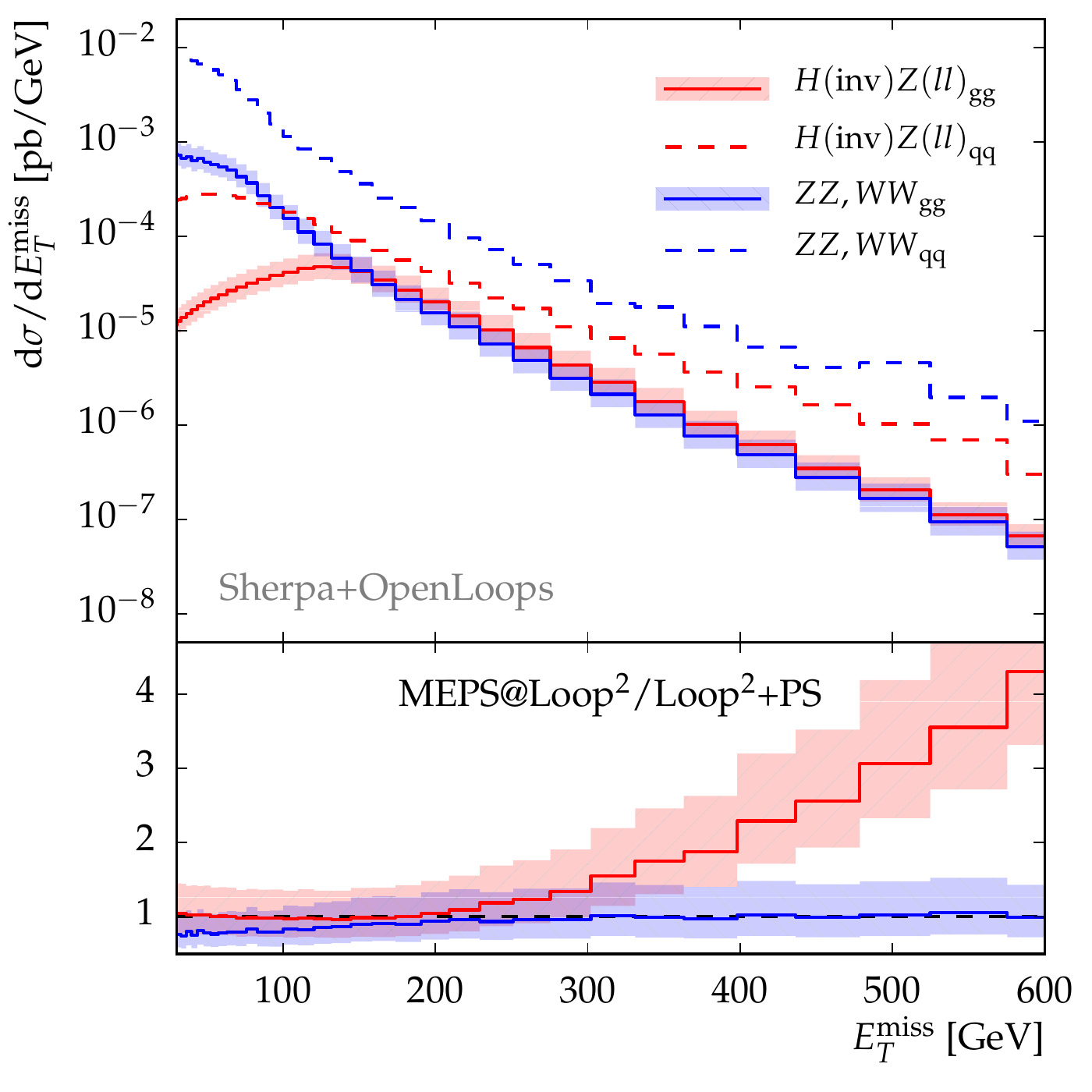}
  \hspace{0.5cm}
  \includegraphics[width=.48\textwidth]{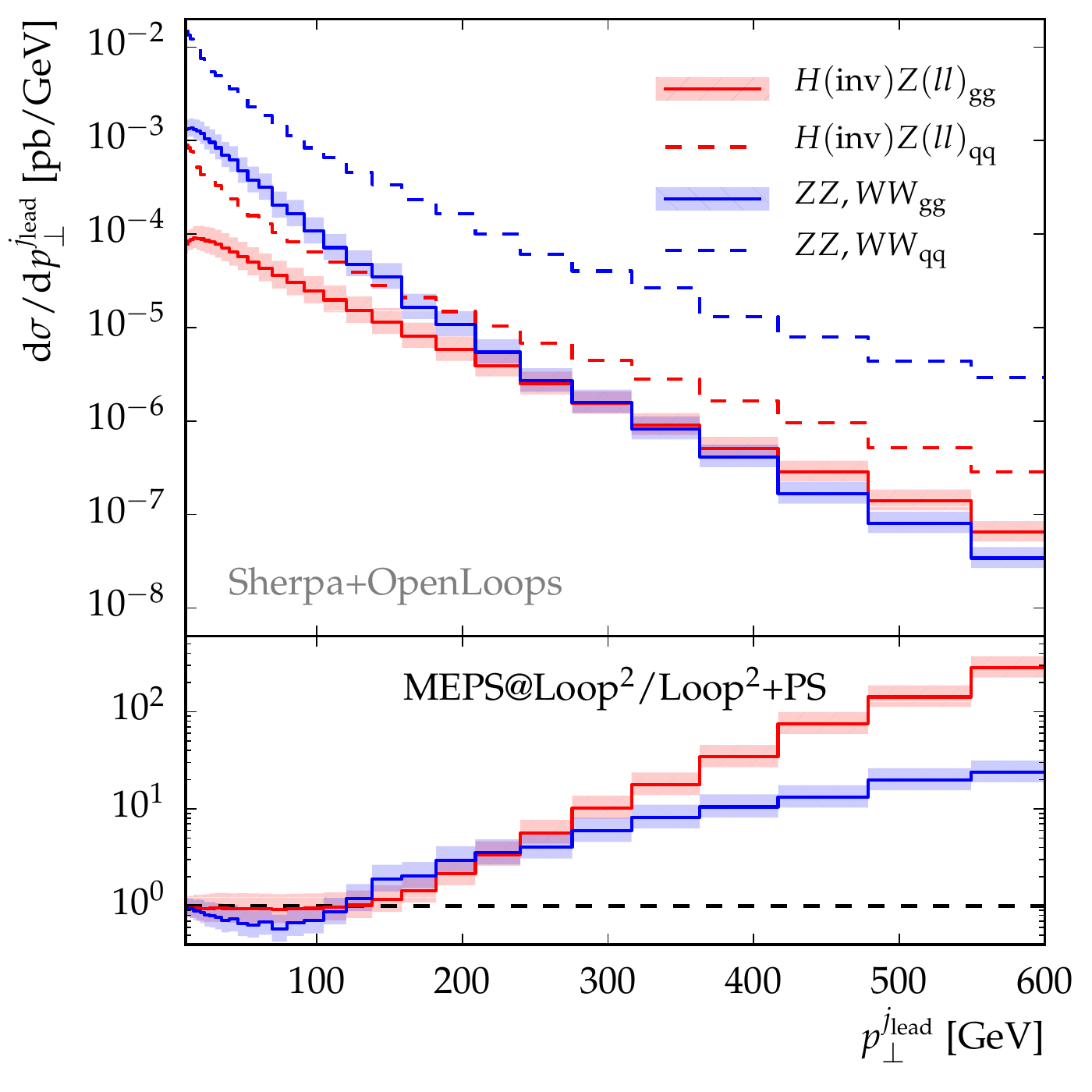}
  \caption{
    Missing energy distribution $\slashed{E}_T$ (left) and leading jet
    transverse momentum $p_T^{j_{lead}}$ distribution (right) for the
    signal $Z(ll)H(\text{inv})$ and background $W^{(*)}W^{(*)},Z^{(*)}Z^{(*)}$
    components assuming ${\mathcal{BR}(H\rightarrow inv)=1}$.
    The signal and background are both decomposed in Drell-Yan like and
    loop-induced components, which are generated by {\sc MEPS@NLO} and
    {\sc MEPS@Loop$^{2}$}, with up to one-jet merged.  The bottom panel
    displays the ratio between the merged sample and the pure LO (i.e.\ box    
    only) plus Parton Shower loop-induced samples for both signal (red) and
    background (blue).}
  \label{fig:merging}
\end{figure*}

Notably, there are two main requirements in the invisible searches that
result in enriching these loop-induced contributions. First, the analysis
usually accounts separately for the zero- and one-jet exclusive bins. This
strategy is very efficient to suppress the initially overwhelming
$t\bar{t}$+jets background, especially in association with a $b$-tagging
veto. Second, it requires large missing energy in the event selection,
usually ${\slashed{E}_T\gtrsim 100}$~GeV. This selection, in particular
depletes the $Z(ll)+$jets background~\cite{inv_exp}. Combining these two
ingredients results in important phenomenological implications.  They are
illustrated in Fig.~\ref{fig:met}, where the missing energy distributions
are displayed for the signal $ZH$ and background $VV'$ in the zero- (left)
and one-jet (right) exclusive bins.

While the GF $VV'$ background presents an almost flat contribution over the
whole missing energy distribution, the $ZH$ signal shows a phenomenologically
relevant threshold at $\slashed{E}_T\sim m_t$. The GF signal is driven by the
heavy flavour quark loops with their characteristic branch cut at
$m_{ZH}\sim 2m_t$ resulting in relevant rates for the boosted regime. On the
other hand, the GF background is dominated by light flavour quark corrections,
without any such structure.  The also present top-quark loop contribution,
which of course also features a branch cut, enhances at the boosted regime
as well.  However, this is a sub-leading correction only when compared to the
other five light flavour quark loops. Hence, no phenomenologically relevant
enhancement is observed in this component at the boosted regime.

The GF signal presents relevant effects that can tantamount to $\sim30\%$
of the signal rate for the zero-jet bin going to up to $\sim50\%$ for the
one-jet bin around the top mass threshold. The GF $VV'$ background present
smaller contributions entailing approximately $\sim10\%$ of its background
rate for the zero-jet and $\sim15\%$ for the one-jet. The larger initial
state colour factor for the GF leads to a higher radiation probability in
comparison to the DY component. {\it I.e.}, the zero-jet sample tends to be
more populated by the DY component and the higher multiplicities receive
larger contributions from the GF. Thus, robust predictions for both the
signal and background samples have to account for the GF component.
We note that the GF signal component also renders important contributions
to the hadronic Higgs decay channel $ZH(b\bar{b})$. In particular, it
leads to phenomenologically relevant modifications on the invariant mass
distribution to the Higgs fat-jet. This phenomenological effect has direct
impact for instance on the bottom Yukawa bounds. See Ref.~\cite{zh_ours}
for more details.

\subsection{Multi-jet Merging}
\label{sec:merging}

The separation of the signal and background in jet bins has become a common
ingredient to many LHC analysis with complex backgrounds. For the invisible
searches, as previously highlighted, this procedure is also customary since
the initially overwhelming $t\bar{t}$+jets background can be set under control
with jet vetoes.

The tool of choice to properly account for the detailed QCD emissions in each
jet sample is multi-jet merging. Our simulation takes into account the
following contributions. The Drell-Yan like signal $Z(ll)H$ and background
$V^{(*)}V'^{(*)}$ are merged up to one jet at NLO precision through the
{\sc MEPS@NLO} algorithm~\cite{mepsnlo}, which can be understood as the
combination of towers of \textsc{MC@NLO} simulations into one inclusive
sample without double counting of extra emissions~\cite{mcatnlo,smcatnlo}.
The respective loop-induced components are generated at LO and also merged
up to one jet, denoted by {\sc MEPS@Loop$^{2}$}~\cite{ckkw}.
\medskip

\begin{figure*}[!t]
  \includegraphics[width=.43\textwidth]{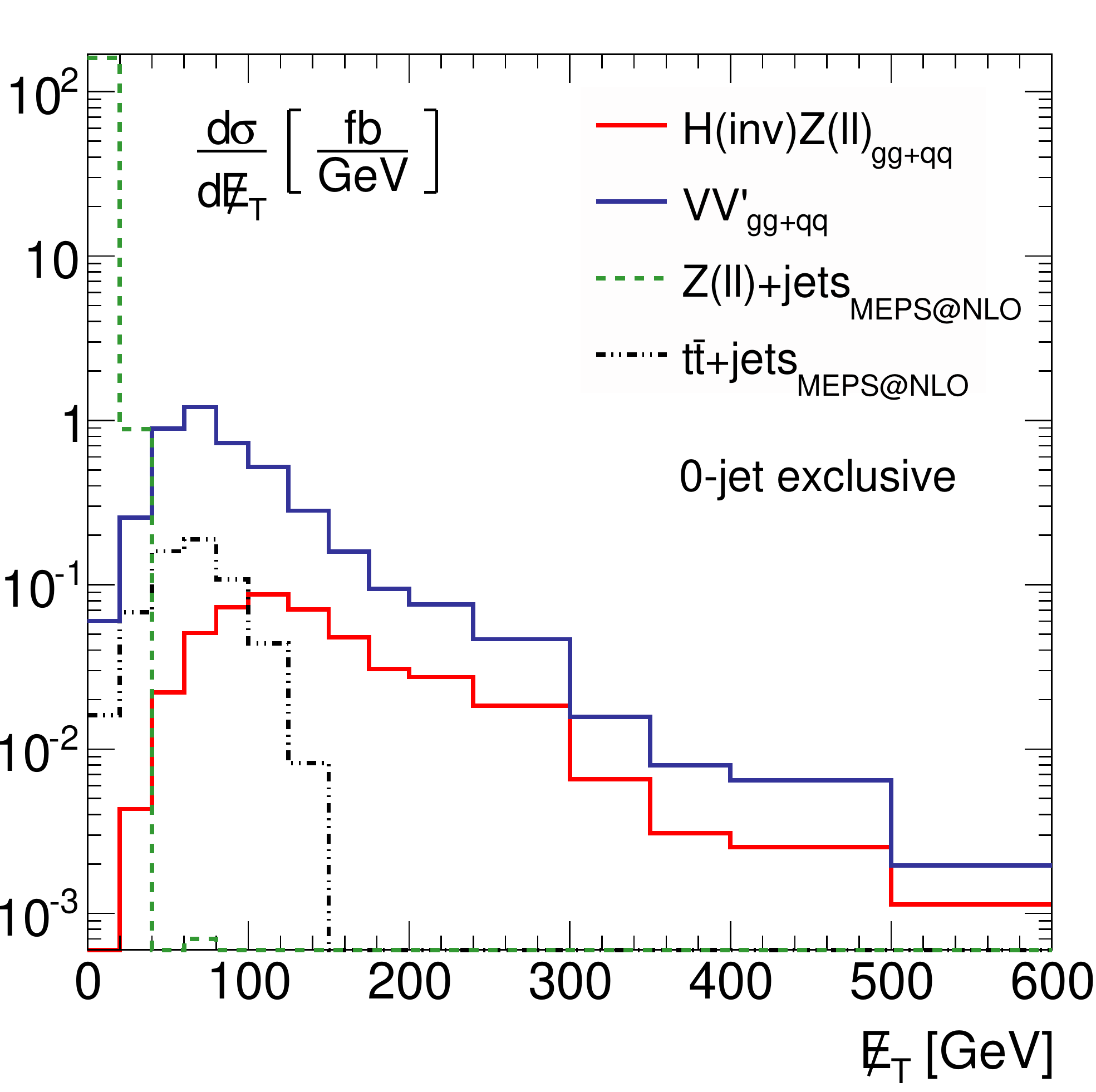}
  \hspace{0.7cm}
  \includegraphics[width=.43\textwidth]{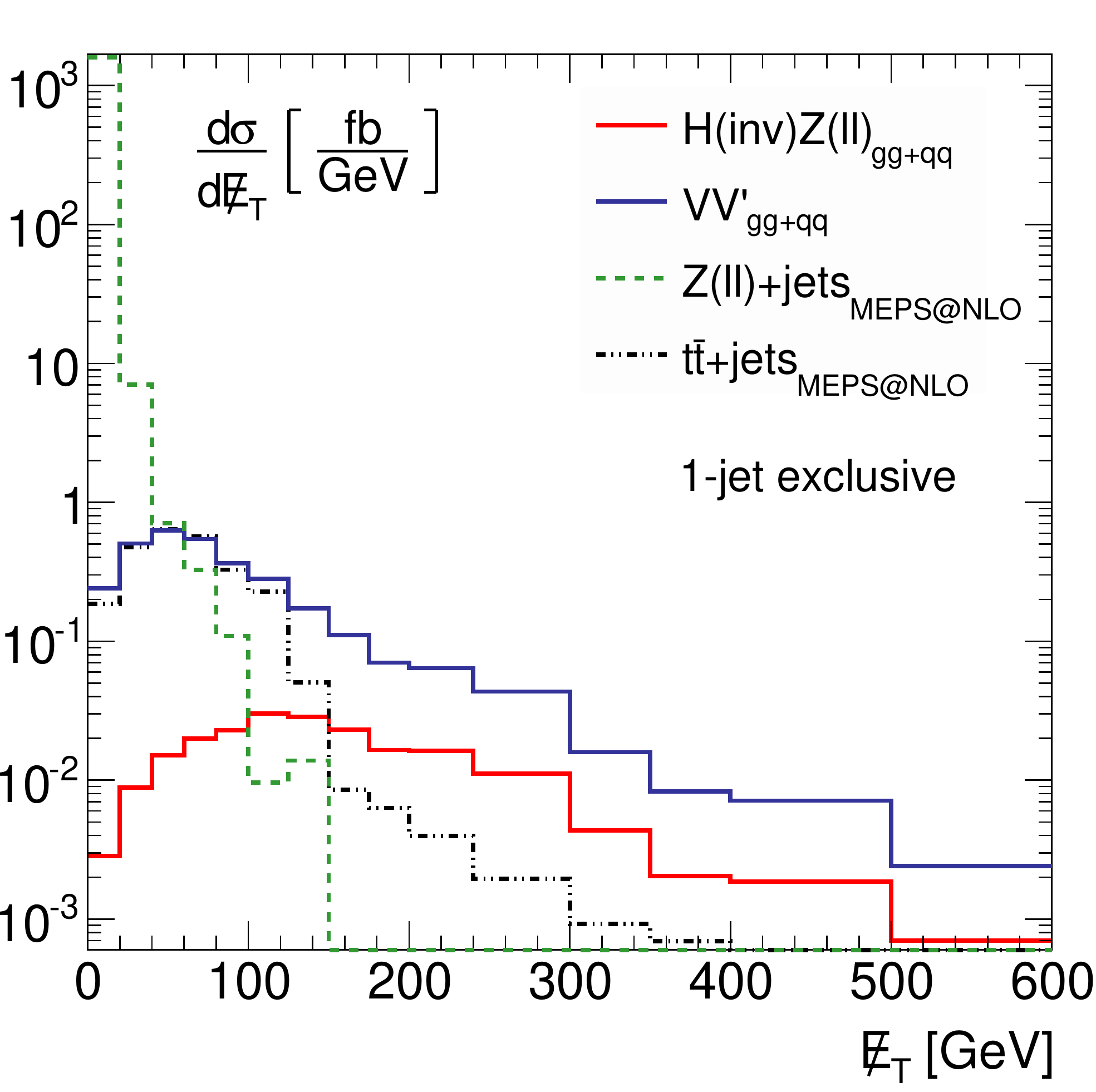}
  \caption{
    Missing transverse energy distribution $\slashed{E}_T$ for the signal
    $H(inv)Z(ll)$ (red) and its major background components $VV'=ZZ,ZW,WW$
    (blue), $Z(ll)+$jets (green) and $t\bar{t}$+jets (black). The zero-jet exclusive 
    bin distributions are shown on the left and the one-jet exclusive on the right.}
  \label{fig:met_sig_back}
\end{figure*}

In Fig.~\ref{fig:merging} (left) the missing energy distribution is displayed
for the signal and $VV'$ background components. We observe that they produce
similar rates either for their DY (dashed) or GF (full) components at the
phenomenologically relevant boosted kinematics regime. In the lower panel, 
the ratio of the multi-jet merged sample ({\sc MEPS@Loop$^{2}$}) to the naive
LO plus Parton Shower ({\sc Loop$^{2}$+PS}) GF samples is shown. The latter
is the approach typically followed in current experimental studies for the
signal predictions~\cite{inv_exp}, where the GF jet emission is based only
on the Parton Shower approximation. Although the background presents a flat
correction profile, the signal displays a major enhancement with the missing
energy distribution for $\slashed{E}_T>m_t$ that can reach up to a factor
of $\mathcal{O}(4)$ at $\slashed{E}_T\sim 600$\,\gev~\cite{zh_ours}. This
effect has a similar origin to the $H$+jets top mass contributions at the
boosted regime~\cite{buschmann2,buschmann1}. The missing transverse energy
can obtain its recoil from jet emissions, which directly probe the loop
structure. This loop factor is dominated by the top quark mass effects in
the signal case. If the $\slashed{E}_T$ exceeds the top mass threshold, the
effects of the loops in the corresponding matrix elements become large. This
feature is not captured in the {\sc Loop$^2$+PS} approximation where the jet
is generated purely through an initial state gluon splitting. Since it is
fundamentally related to the top-quark contributions, the signal presents
a large correction while the background, that is predominantly generated by
light quark loops, do not produce any appreciable change.
\medskip

Merging effects are even larger when considering jet observables. Without
merging, any extra parton level QCD radiation is generated only by the
parton shower and hard jets are, correspondingly, not appropriately described.
In Fig.~\ref{fig:merging} (right), we present the transverse momentum
distribution of the hardest jet for both signal and background contributions.
Large discrepancies show up when comparing the spectra obtained from a
{\sc Loop$^2$+PS}-type simulation with the ones obtained from the
corresponding merged samples. In the region of large transverse momenta,
the {\sc Loop$^2$+PS} prediction underestimates the spectrum by orders of
magnitude. For example, a relative factor of $\mathcal{O}(10)$ is observed
for $p_T^{j_{lead}}\sim 300$~GeV which is even further enhanced at higher
energies. In this regime, the soft/collinear approximation inherent to the
parton shower fails. By merging matrix elements with one additional jet into
the sample, we recover the corresponding fixed-order matrix element accuracy
that is required for an appropriate description in this regime. The signal
and background merging corrections present a similar pattern below the top
threshold. Above this threshold, however, again large corrections for the
signal sample are observed.  This essentially recovers the results already
present in Fig.~\ref{fig:merging} (left), where the effects in the
signal sample where much larger than in the background for this 
regime.\footnote{See Appendix~\ref{sec:appendix} for further details on the
multi-jet merging for loop-induced processes.}
\medskip

In conclusion, any robust theoretical description with the usual separation
in jet bins and the boosted kinematics selections requires the inclusion of
the loop-induced components and of multi-jet merging algorithms -- for both
the signal and the background.

\section{Constraints on Invisible decays}
\label{sec:inv}

In this section, the constraining power of the Higgs-Strahlung $Z(ll)H$
channel to the branching ratio of invisible decays of the Higgs boson,
$\mathcal{BR}(H\rightarrow inv)$ at the $\sqrt{s}=13$~TeV LHC is analysed.
The major backgrounds for this process are di-boson pair
${V^{(*)}V'^{(*)}=WW,WZ,ZZ}$, top pair $t\bar{t}+$jets and $Z+$jets production.

The Monte Carlo studies in this publication are performed with
\textsc{Sherpa+OpenLoops}~\cite{sherpa,openloops,collier}. The DY $Z(ll)H$,
DY $VV'$, $t\bar{t}+$jets and $Z+$jets samples are generated with the
\textsc{MEPS@NLO} algorithm~\cite{mepsnlo}, with up to one extra jet
at NLO (QCD) accuracy for all processes apart fron $Z+$jets, where up to
two jets have been treated at NLO. The loop--\-induced GF components are
generated with the \textsc{MEPS@Loop$^2$} algorithm~\cite{ckkw}. These samples
are again merged up to one extra jet, this time at LO.  Finite width effects
and spin correlations from the leptonic vector boson decays are fully
accounted for in the simulation. Hadronisation and underlying event effects
are also included.
\medskip

In our event analysis, we require two isolated, same-flavour, opposite-sign
leptons with $p_{Tl}>20$~GeV and ${|\eta_l|<2.5}$. The lepton isolation
criterion demands less than 20\% of hadronic activity in a radius of
$R=0.2$ around the lepton. The invariant mass of the di-lepton system
$m_{ll}$ is required to fall into the $Z$-boson mass window
${|m_{ll}-m_Z|<15}$~GeV.  Jets are defined with the anti-$k_T$ jet algorithm
with radius $R=0.4$, ${p_{Tj}>30}$~GeV and $|\eta_j|<5$ using the
\textsc{fastjet} package~\cite{fastjet}. To suppress the initially
overwhelming $t\bar{t}$ background, b-tagged jets are vetoed, assuming a 
70\% b-tagging efficiency and 1\% mistagging rate~\cite{btagging,btagging_ours}.
Throughout a Gaussian smearing of $\Delta\slashed{E}_T=20$~GeV is applied to 
the missing energy vector.

Since most of the signal sensitivity is in the boosted regime
${\slashed{E}_T>100}$~GeV, where the $Z$ boson decays are produced with small
opening angles, we require an extra event selection on their azimuthal angle
${\Delta \phi(l,l)<1.7}$. This selection is efficient to further suppress
in particular the $t\bar{t}$ and $W(l\nu)Z(ll)$ backgrounds.

Binning in jet multiplicities has been established as an efficient tool to
further control the $t\bar{t}$ and other backgrounds. In our analysis, and
following standard procedures in similar experimental analyses, the full event
sample is divided into zero- and one-jet exclusive subsamples.  In
Fig.~\ref{fig:met_sig_back}, the missing transverse energy distribution,
$\slashed{E}_T$, is depicted for the signal and background components for
the zero- (left panel) and one-jet (right panel) exclusive samples. The
combination of jet vetoes and large missing energy selections tame both the
$Z$+jets and $t\bar{t}$ backgrounds. Nonetheless, these selections do not
result in major extra gains with respect to $VV'$ background. The structural
similarities of this background with respect to the signal, as discussed in
Sec.~\ref{sec:zh_loop}, result in comparable contributions throughout the
whole missing energy distribution profile.\medskip

\begin{figure}[!b]
  \includegraphics[width=.5\textwidth]{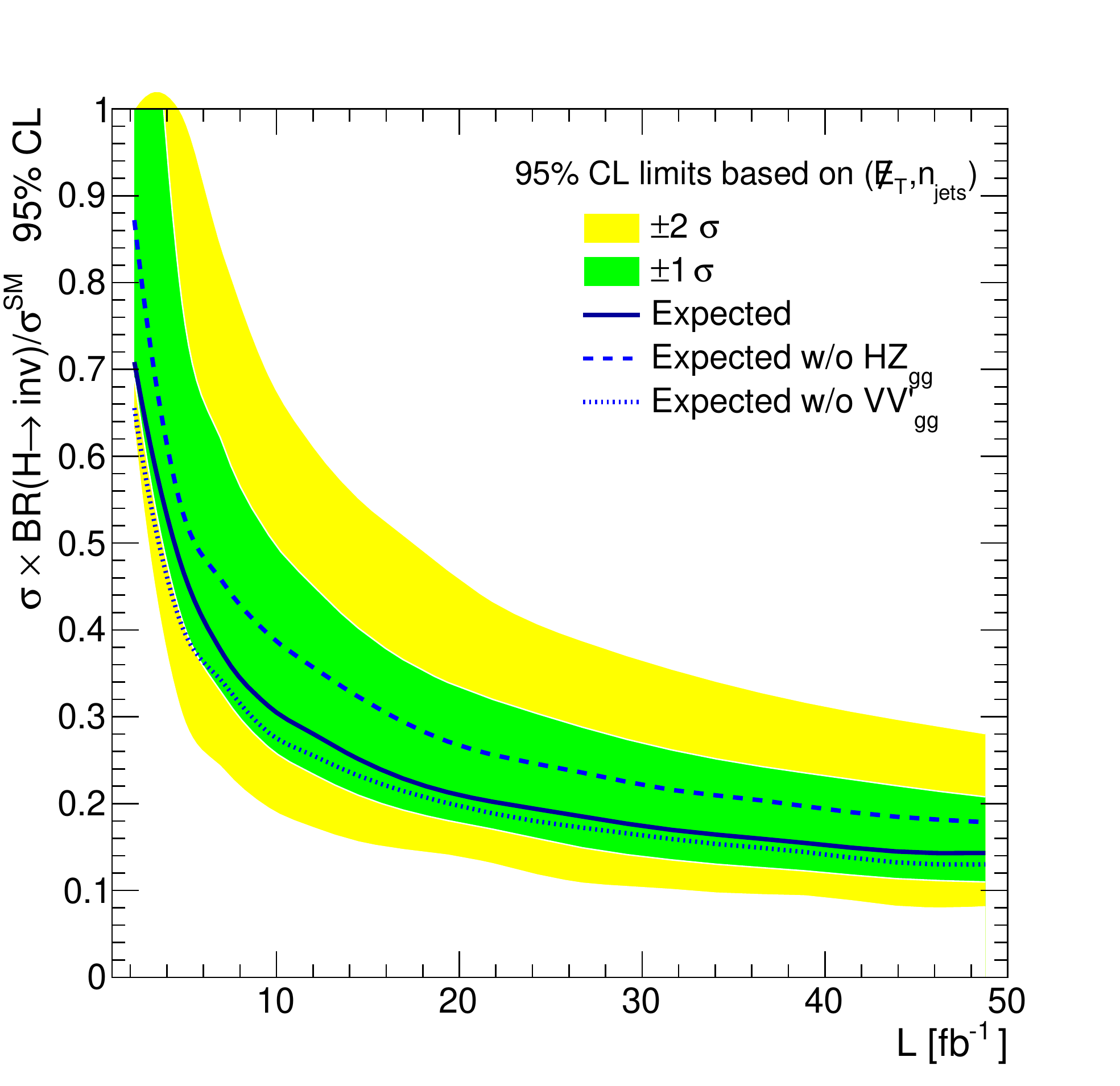}
  \caption{
    Expected 95\% CL upper limit on the $Z(ll)H$ production times the
    invisible Higgs branching ratio ${\mathcal{BR}(H\rightarrow inv)}$
    normalised to the SM production cross section. For comparison, also
    the expected upper limits are shown when not accounting for the
    loop-induced $V^{(*)}V'^{(*)}_{gg}$ background (dotted line) and when the
    loop-induced $HZ_{gg}$ is neglected (dashed line).}
  \label{fig:cls}
\end{figure}

To estimate the constraining power to the Higgs to invisible branching
ratio ${\mathcal{BR}(H\rightarrow inv)}$ from this study, we perform a
two-dimensional binned log-likelihood test for the missing energy
distribution $\slashed{E}_T$ vs.\ the number of jets $n_{jets}=0,\,1$.  This 
procedure exploits the distribution shapes of both panels displayed in
Fig.~\ref{fig:met_sig_back} by invoking the CL$_s$ method~\cite{Junk:1999kv}.
We show in Fig.~\ref{fig:cls} the 95\% CL upper limit to the invisible Higgs
boson branching ratio from the Run II LHC.  It is possible to bound the
branching ratio to ${\mathcal{BR}(H\rightarrow inv)} \lesssim 0.3$ with only
$10~fb^{-1}$ integrated luminosity.  To allow a direct appreciation of the
importance of the various contributions, in the same figure also the
resulting bound is shown, when neglecting separately the GF $VV'$ background
and GF $HZ$: By neglecting the loop--\-induced background component overly
constraining limits would be produced, differing by more than one standard
deviation from the correctly expected bound, depending on the luminosity.
On the other hand, if the GF signal component was neglected as well, the
simulation would present weaker bounds than the more precise prediction that
accounts for all the components.  At $10~fb^{-1}$ this would result in
shifting the correct bound from ${\mathcal{BR}(H\rightarrow inv)}\lesssim0.3$
to approximately 0.4. 

It is worth stressing that this analysis provides only an upper bound.  Further
improvements can be obtained, for instance, by extensive use of Multivariate 
Analysis (MVA) techniques, combining the distributions discussed here with other
significant distributions.

\section{Constraints on Dark Matter Simplified Models}
\label{sec:DM}

Searches for Beyond the Standard Model (BSM) physics, where the SM
degrees of freedom and the new BSM states are separated by a large energy
gap, are often performed in an Effective Field Theory (EFT) approach.  This
is also true for DM searches at the LHC.  However, usually these searches
require large missing energy selections that render the EFT approach invalid
for a significant range of the parameter space~\cite{Fox:2011fx,
  Shoemaker:2011vi,Weiner:2012cb,Busoni:2013lha,Buchmueller:2013dya,
  Buchmueller:2014yoa,Busoni:2014sya,Busoni:2014haa}.  Instead of resorting
to UV-complete theories, losing the model independency of the derived
constraints, a set of {\it Simplified Models} was constructed where new
particles mediating the interactions between the {\it visible} and the
{\it dark} sectors~\cite{Alwall:2008ag,Alves:2011wf,Goodman:2011jq,
  Jacques:2015zha,Godbole:2015gma,Abdallah:2015ter, Abercrombie:2015wmb}
can be directly produced at colliders. In the present section, we focus on
a class of such Simplified Models with either new scalar or pseudo-scalar
mediators~\cite{Buckley:2014fba,ttDM_CP,Haisch:2012kf,Haisch:2013ata,
  Haisch:2013fla,Crivellin:2014qxa,Englert:2016joy,Ghorbani:2014qpa,
  Backovic:2015soa,Harris:2014hga,Mattelaer:2015haa,Haisch:2015ioa,
  Berlin:2015wwa}.  \medskip
  
We assume a Dirac fermion DM $\chi$ that can be produced
through the decay of either a scalar $\phi$ or a pseudo-scalar $A$ mediator,
which are produced by couplings to SM fermions $f$.  We also assume that the
interaction respects Minimal Flavour Violation, where the couplings are assumed to be
proportional to the Higgs Yukawa interactions.  Under these assumptions,
flavour constraints are avoided, and the Lagrangian for the interaction terms
for the mediator to SM fermions and DM is given by
\begin{alignat}{5}
  \mathcal{L}\supset & -\sum_{\text{f}}\frac{y_f}{\sqrt{2}}
  \left(g_{v}^{\phi}\phi\bar{f}f+ig_{v}^{A}A\bar{f}\gamma_5 f\right)\notag\\
  & -g_\chi^{\phi}\phi\bar{\chi}\chi - i g_\chi^{A}A \bar{\chi}\gamma_5\chi\,,
  \label{eq:Lphi}
\end{alignat}
where $g_\chi$ is the DM--mediator coupling and $g_v$ the universal SM
fermion--mediator strength.
\medskip

\begin{figure}[!t]
  \centering
  \begin{tikzpicture}[]
 \node [on grid] (gu) at (-1.5,+0.5) {$g$};
 \node [on grid] (gd) at (-1.5,-0.5) {$g$};

 \node [vertex] (tul) at (-0.5,+0.5) {};
 \node [vertex] (tdl) at (-0.5,-0.5) {};
 \node [vertex] (tur) at (+0.5,+0.5) {};
 \node [vertex] (tdr) at (+0.5,-0.5) {};
 \draw [gluon]  (gu)  -- (tul);
 \draw [gluon]  (gd)  -- (tdl);
 \draw [dirac]  (tul) -- (tdl) node [midway, right] {$t,b$};
 \draw [dirac]  (tdl) -- (tdr);
 \draw [dirac]  (tdr) -- (tur);
 \draw [dirac]  (tur) -- (tul);

 \node [vertex] (zu)  at ($(tur)+(0.9,+0.2)$) {};
 \node [      ] (zd)  at ($(tdr)+(1.2,-0.2)$) {$\phi/A$};
 \draw [vector] (tur) -- (zu) node [midway, above] {\small $Z$};
 \draw [scalar] (tdr) -- (zd);

 \node [on grid] (lu1)  at ($(zu)+(0.75,+0.5)$) {$l^+$};
 \node [on grid] (lu2)  at ($(zu)+(0.75,-0.5)$) {$l^-$};
 \draw [dirac]   (lu1) -- (zu);
 \draw [dirac]   (zu) -- (lu2);

 \begin{scope}[shift={(4,0)}]
   \node [on grid] (gu) at (-1.5,+0.5) {$g$};
   \node [on grid] (gd) at (-1.5,-0.5) {$g$};

   \node [vertex] (tul) at (-0.5,+0.5) {};
   \node [vertex] (tdl) at (-0.5,-0.5) {};
   \node [vertex] (tur) at (+0.5,+0.6) {};
   \node [vertex] (tr)  at (+0.7,+0.0) {};
   \node [vertex] (tdr) at (+0.5,-0.6) {};
   \draw [gluon]  (gu)  -- (tul);
   \draw [gluon]  (gd)  -- (tdl);
   \draw [dirac]  (tul) -- (tdl) node [midway, right] {$t,b$};
   \draw [dirac]  (tdl) -- (tdr);
   \draw [dirac]  (tdr) -- (tr);
   \draw [dirac]  (tr)  -- (tur);
   \draw [dirac]  (tur) -- (tul);

   \node [vertex] (zu)  at ($(tur)+(0.9,+0.4)$) {};
   \node [      ] (zd)  at ($(tdr)+(1.2,-0.4)$) {$\phi/A$};
   \draw [vector] (tur) -- (zu) node [midway, above left] {\small $Z/\gamma$};
   \draw [scalar] (tdr) -- (zd);

   \node [on grid] (lu1)  at ($(zu)+(0.75,+0.5)$) {$l^+$};
   \node [on grid] (lu2)  at ($(zu)+(0.75,-0.5)$) {$l^-$};
   \draw [dirac]   (lu1) -- (zu);
   \draw [dirac]   (zu)  -- (lu2);

   \draw [gluon] (tr) --++ (0.7,0.0) node [right] {$g$};
 \end{scope}
\end{tikzpicture}
  \caption{
    Representative loop-induced Feynman diagrams contributing to the signal
    $l^+l^-\phi/A$ (left) and $l^+l^-\phi/Aj$ (right)}
  \label{fig:diags_DM}
\end{figure}
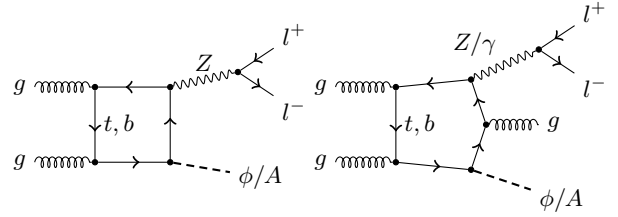

\begin{figure}[!b]
  \includegraphics[width=.48\textwidth]{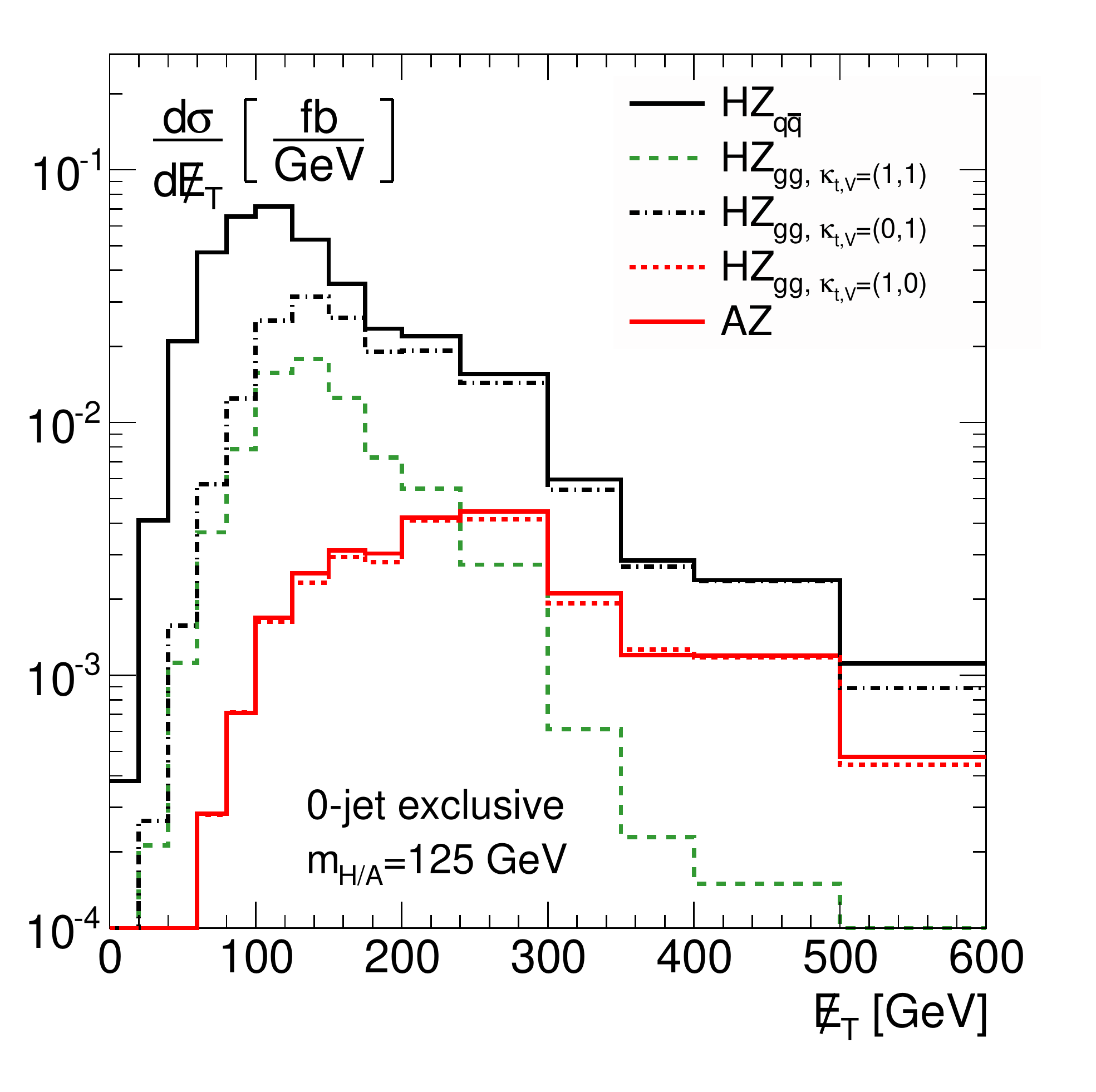}
  \caption{
    Transverse missing energy distribution $\slashed{E}_T$ for
    ${pp\rightarrow Z(ll)H(inv)}$.  The following components are shown:
    DY (black full), GF (green dashed), GF with the scalar--vector boson
    interactions switched off ${\kappa_{t,V}=(1,0)}$ (red dashed) and GF with
    the Yukawa interactions switched off ${\kappa_{t,V}=(0,1)}$ (black
    dotted-dashed). Notice that the GF curve with ${\kappa_{t,V}=(1,0)}$
    corresponds to our scalar simplified model.  For comparison, also
    the pseudo-scalar channel ${pp\rightarrow Z(ll)A(inv)}$ (red full) is
    added.  In this plot $m_{H/A}=125$~GeV and
    $\mathcal{BR}(H/A\rightarrow inv)=1$.
    The DY and GF samples are merged up to one jet respectively 
    with {\sc MEPS@NLO} and {\sc MEPS@Loop$^2$} technology.}
  \label{fig:met_DM}
\end{figure}

\begin{figure*}[!t]
  \includegraphics[width=.49\textwidth]{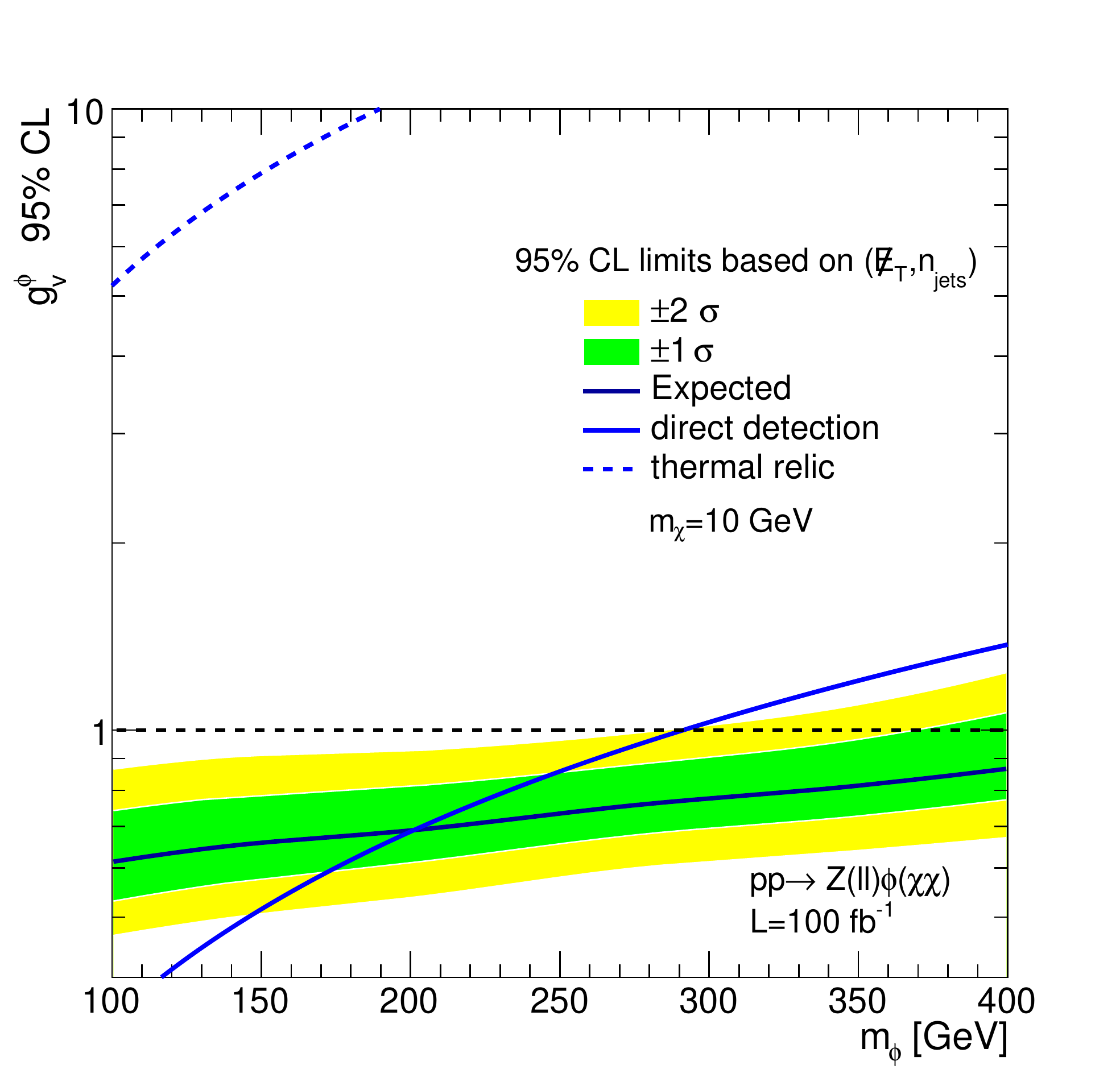}
  \includegraphics[width=.49\textwidth]{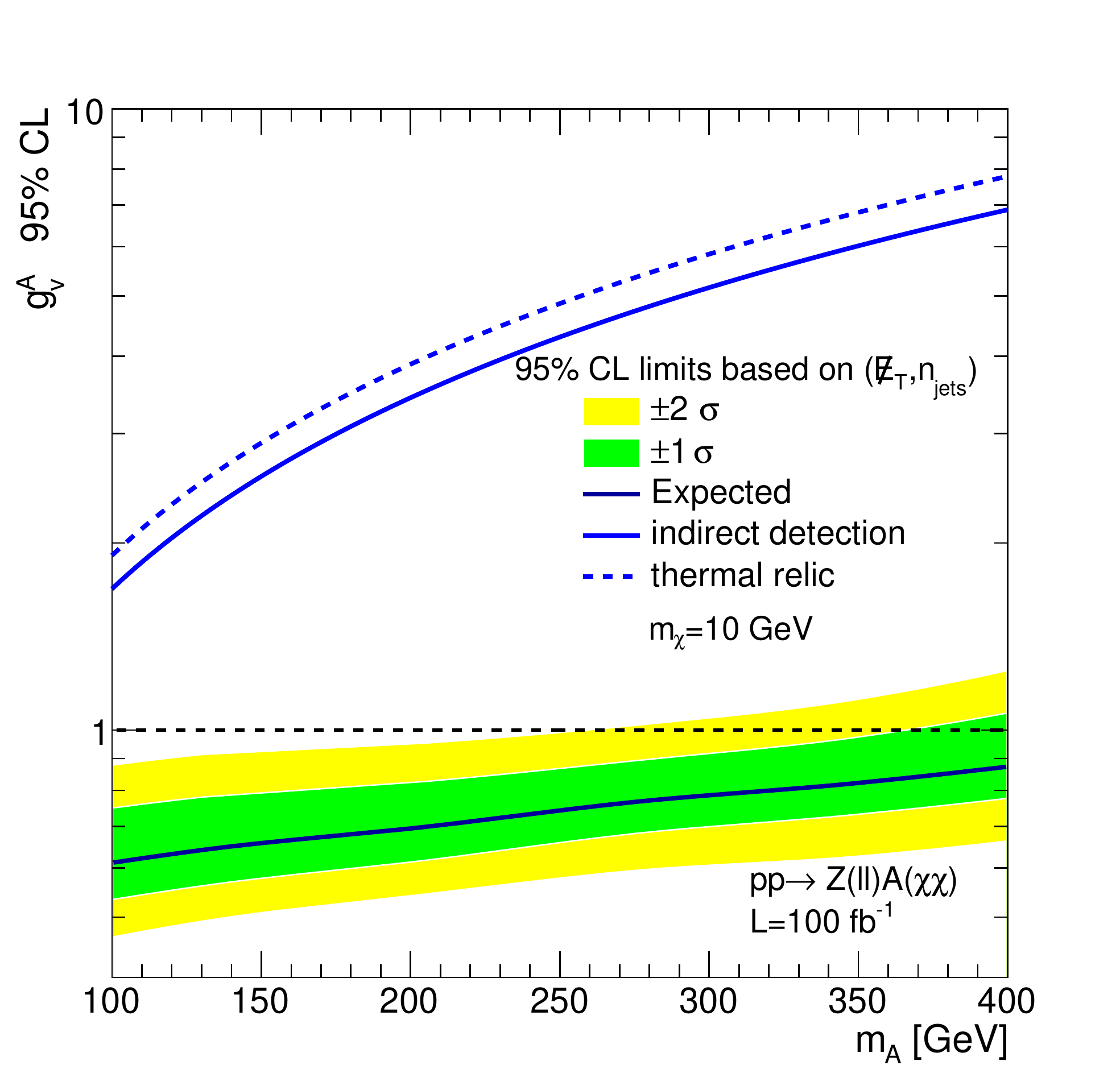}
  \caption{
    Expected 95\% CL upper limit on the mediator coupling to fermions $g_v$
    as a function of the mediator mass $m_{\phi/A}$ for scalars (left) and
    pseudo-scalars (right). The binned log-likelihood test is based on the two
    dimensional distributions $(\slashed{E}_T,n_{jets})$ for $100~fb^{-1}$ of LHC
    data assuming $\mathcal{BR}(\phi /A\rightarrow\chi\chi)=1$. In addition
    the 95\% CL upper limits on scalar mediators from \textsc{LUX}
    direct detection bounds~\cite{lux} and on pseudo-scalars from the
    \textsc{Fermi-LAT} dwarf galaxy search~\cite{fermilat} are included. The
    required $g_v$ value for thermal relic is also shown in both cases.}
  \label{fig:cls_DM}
\end{figure*}

Such interactions can be probed by multiple channels at the LHC, producing
interesting searches for example with missing energy plus top or bottom
quarks or jets~\cite{Buckley:2014fba}.  Furthermore, the CP nature of the
mediator can also be directly probed through spin correlations in the DM
production associated with tops~\cite{ttDM_CP,Buckley:2015vsa}.  Besides
these well studied signatures, the interactions also result in di-lepton
plus missing energy signatures from the loop-induced $Z(ll)\phi(\chi\chi)$
or $Z(ll)A(\chi\chi)$ channels, with sizable event yields.  In the following
we derive the complementary LHC bounds that result from these new channels.
\medskip

In Fig.~\ref{fig:diags_DM} a representative set of the Feynman diagrams
contributing to the DM signal production is shown. While the box-like diagrams
only contribute with the Z boson axial-vector coupling as a consequence of
Furry's theorem, the pentagon diagrams can have both vector and axial-vector
contributions, allowing particularly $Z$ and photon interference terms. These
statements hold for both the scalar and pseudo-scalar hypotheses, implying 
similar total event rates for both scenarios.

The simplified DM signal for the case of a scalar mediator can be obtained
in a straightforward way from the GF $ZH$ production in the SM, by simply
turning off the $HZZ$ EW coupling, $\kappa_{t,V}=(1,0)$.  For comparison,
the $pp\rightarrow Z(ll)H(inv)$ is decomposed in Fig.~\ref{fig:met_DM} into a
DY and a GF$_{\kappa_{t,V}=(1,1)}$ component, as discussed in some detail in the
previous sections.  The GF is further decomposed by separately switching off
the fermion couplings, GF$_{\kappa_{t,V}=(0,1)}$, and the EW couplings,
GF$_{\kappa_{t,V}=(1,0)}$.  The latter, GF$_{\kappa_{t,V}=(1,0)}$ (i.e., the scalar
DM simplified model), presents enhanced rates in the boosted regime that
are comparable to the DY component.  To inspect differences between
scenarios with scalar and pseudo-scalar mediators, the $\slashed{E}_T$
distribution in ${pp\rightarrow Z(ll)A(inv)}$ is also added.  It is
very similar in both rate and shape to the loop-induced $Z(ll)H(inv)$
contribution, differing by less than 10\% over the full missing energy
distribution.

We reproduce the analysis strategy presented in the previous section and
perform a two-dimensional binned log-likelihood analysis based on the
$\slashed{E}_T$ and $n_{jets}$ distributions.  In Fig.~\ref{fig:cls_DM}, the
expected 95\% CL upper limit on the mediator coupling to fermions $g_v$
is depicted as a function of the mediator mass $m_{\phi/A}$ for scalars (left)
and pseudo-scalars (right).  Assuming $100~fb^{-1}$ of data and
$\mathcal{BR}(\phi /A\rightarrow \chi \chi)=1$, we can bound the signal over
the full considered mass range $100~\text{GeV}<m_{\phi/A}<400$~GeV to $g_v<1$.
Furthermore, the bounds for the two Simplified Models are very similar as
a result of their comparable distribution profiles which differ by only a
few percent.  Fig.~\ref{fig:cls_DM} also shows the 95\% CL non-collider
experimental bounds for DM direct and indirect detection, as well as for
thermal relic abundance cross section bounds~\cite{Buckley:2014fba,ttDM_CP}.
While the pseudo-scalar mediator does not induce a momentum--\-independent
scattering cross-section with nuclei, and so does not display relevant
limits from direct detection experiments, the scalar mediator scenario
generates a spin-independent rate resulting in strong bounds from several
experiments. The most stringent ones come from \textsc{LUX}~\cite{lux}
for large DM masses $m_\chi>6$~GeV and from \textsc{CDMS-lite}~\cite{cdms}
for lower DM masses. The indirect detection bounds are obtained from the
$b\bar{b}$-channel using the Fermi Large Area Telescope (LAT) data on dwarf
galaxies~\cite{fermilat}. This bound is only relevant to the pseudo-scalar
mediators because scalar mediators present velocity suppressed thermal
cross-sections and DM in the present Universe is moving much slower than the
speed of light ($T\ll m_\chi$). Lastly, both mediator models display relevant
bounds from thermal relic abundance,  assuming
${\langle \sigma v\rangle=3 \times 10^{-26}}$~cm$^3$/s in the early universe
and $g_v=g_\chi$ to make the comparisons to collider bounds more
straightforward.  Under the presented assumptions, the LHC can provide
stronger constraints than the non-collider limits for almost the entire
considered mediator mass region.  It only presents weaker constraints for
lower scalar mediator masses ${m_{\phi}<200}$~GeV, where the direct detection
bounds are more relevant.  Importantly, pseudo-scalar mediator scenarios are
much more challenging for non-collider experiments and the LHC bounds
become even more important.

It must be stressed at this point, however, that no single result quoted
here, from collider and non-collider experiments, should be taken as the
final word. The analyses discussed here approach the same problem from
different angles and with different assumptions. For instance, these bounds
can be significantly changed if more particles are present in the spectrum
beyond our  benchmark scenario. In this sense, the presented limits should 
be seen more as a guide that allows us to focus on particular parameter space 
regions with the correspondent experimental data.

\section{Summary}
\label{sec:summary}

In this publication a state of the art analysis for the searches for invisible
decays of Higgs bosons in its $Z$-associated production channel has been
performed, focusing on $\ell\ell+\slashed{E}_T$ final states.  The importance
of the loop--\-induced contributions to both the signal $Z(ll)H(inv)$ and the
backgrounds $WW$ and $ZZ$ has been discussed in detail for the first time,
taking into account the efect of multi-jet merging technology.  Both
contributions lead to relevant changes on important distributions and
therefore effect changes in the $\textsc{CLs}$ bounds that can go beyond the
one-$\sigma$ uncertainty, depending on the collider luminosity.

The two-dimensional $(\slashed{E}_T,n_{jets})$ binned Log-likelihood analysis
in particular shows that the invisible Higgs branching ratio can be bound to
$\mathcal{BR}(H\rightarrow inv)<0.15$ with $\mathcal{L}=50~fb^{-1}$ of data at
the LHC 13~TeV.  In our analysis we confirm that the separation in zero and
one jet bins is fundamental to maximise the control over the otherwise
dominant background $t\bar{t}$+jets.  We also show that multi-jet merging
techniques are fundamental for this analysis, and especially when performing
the separation in jet bins.

We extended the analysis to a set of Simplified Models of Dark Matter which
connect the visible and invisible sectors through scalar or pseudo-scalar
mediators.  We find that the $Z\phi(A)$ channel provides relevant bounds to
this type of models being able to probe the mediator masses in the range of
${100~\text{GeV}<m_{\phi/A}=400}$~GeV with $100~fb^{-1}$.

\begin{appendix}
\section{Merging for loop-induced Processes: validation}
\label{sec:appendix}

\begin{figure}[b!]
  \includegraphics[width=.45\textwidth]{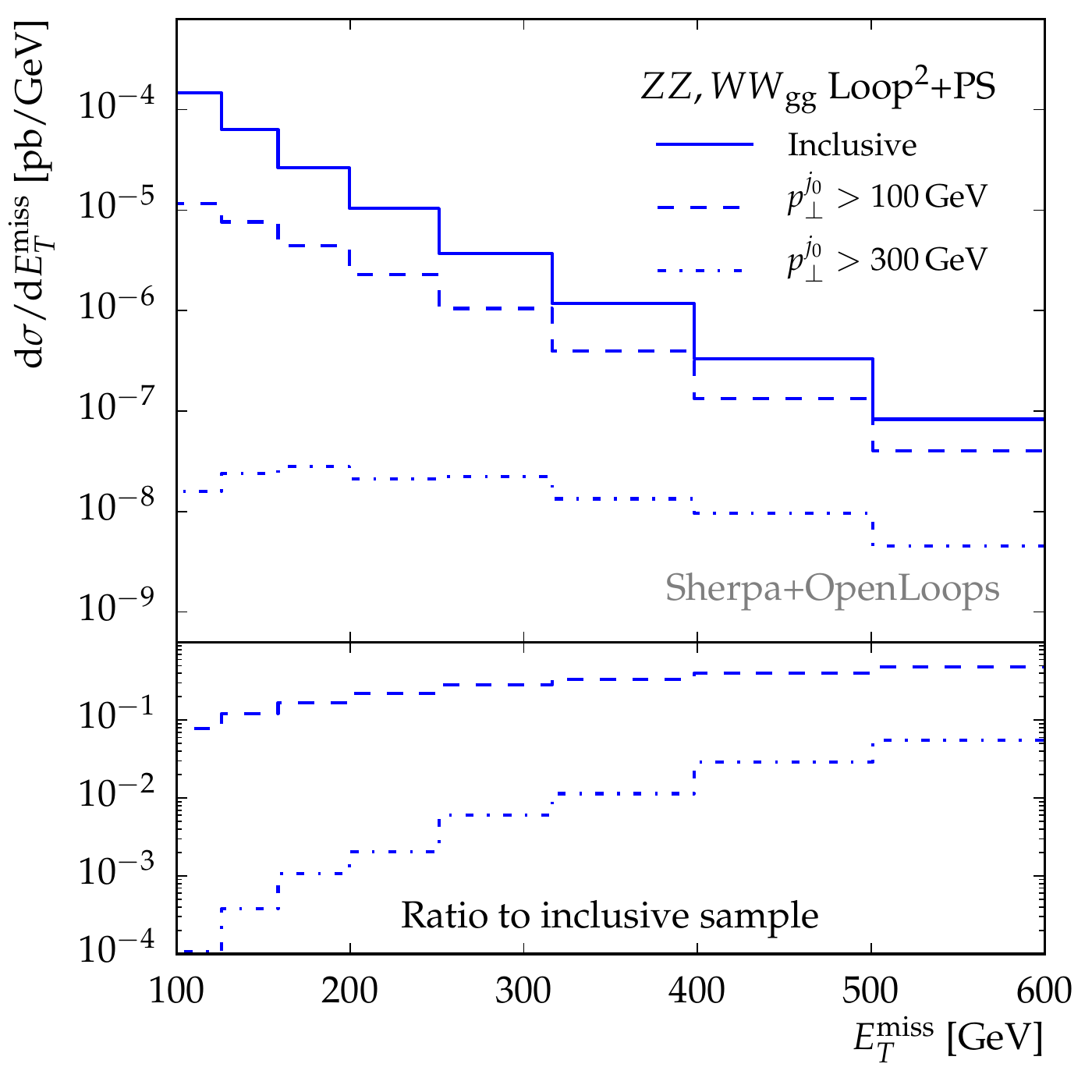}
  \includegraphics[width=.45\textwidth]{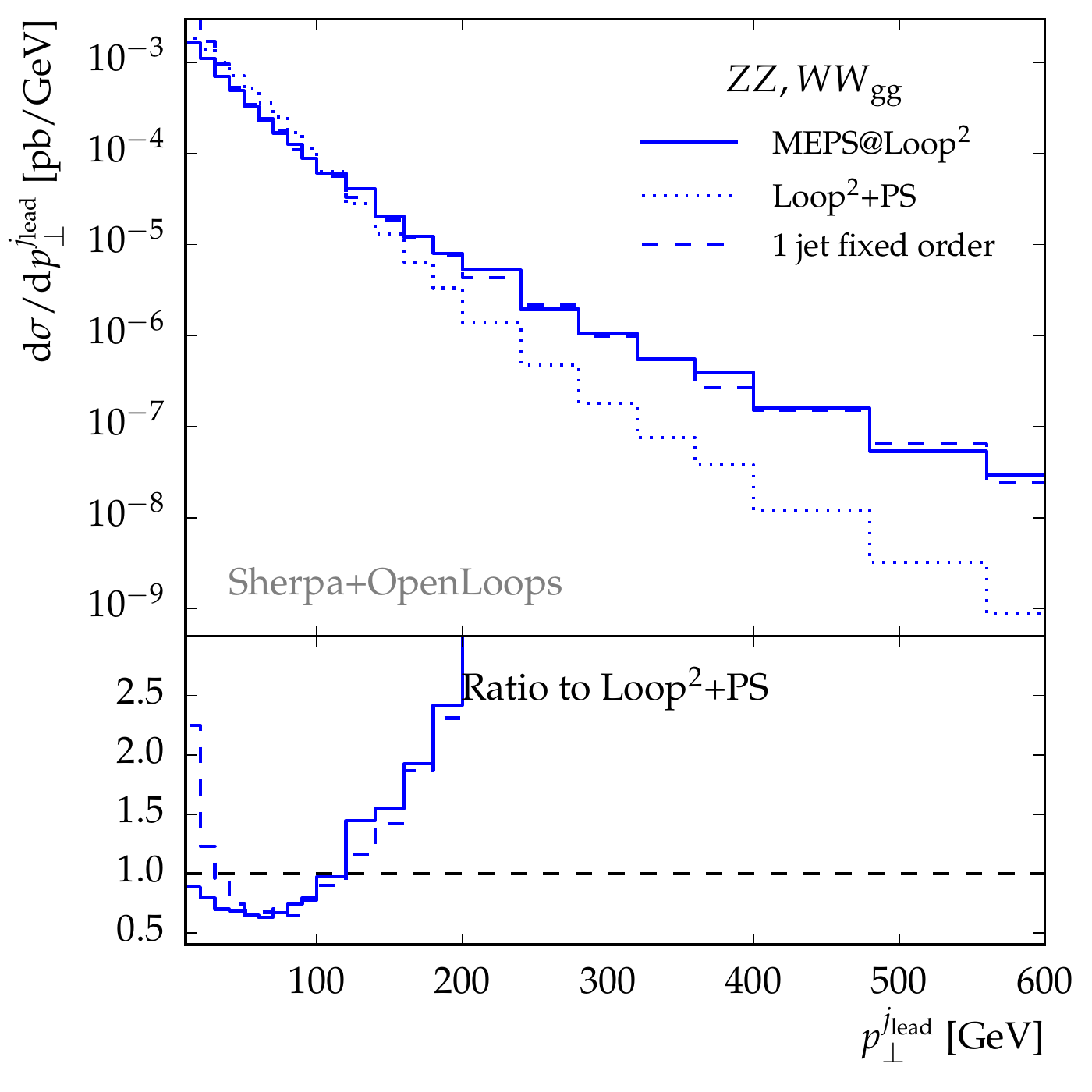}
  \caption{Transverse missing energy $E_T^\mathrm{miss}$  distribution (top)
   and leading jet  transverse momentum $p_T^{j_\mathrm{lead}}$
    spectra (bottom) for the loop-induced component of the
    $ZZ,WW$ process. In the bottom panel, we compare the fixed-order 
    $(ZZj,WWj)_{gg}$ result to the prediction obtained from a multi-jet merging setup.}
  \label{fig:merging_val}
\end{figure}

In Fig.~\ref{fig:met_sig_back}  (right panel), we illustrate how
matrix element corrections included through the merging algorithm
significantly affect the transverse momentum spectrum for high energy jets, 
$p_T^{j_\mathrm{lead}}>100$~GeV. This is an expected feature since the Parton Shower
cannot appropriately  fill phase space regions  where the jet transverse
momentum significantly exceeds the mass of the produced electroweak
final state. Therefore, we observe a strong enhancement in the
$p_T^{j_\mathrm{lead}}$ tail  for both the signal and the background 
processes when employing multi-jet merging.

However, for the background,  this enhancement does not result in an enhancement for
$E_T^\mathrm{miss}$ distribution. Since a high-$p_T$ jet must, to some degree, recoil
against the neutrinos in this process, the lack of an enhancement
in the $E_T^\mathrm{miss}$ distribution is rather surprising.  It can, however,
be tracked down to two circumstances: Firstly, the relative contributions at the 
$E_T^\mathrm{miss}$ tail from large $p_T^{j_\mathrm{lead}}$ configurations are moderate. 
This is explicitly shown in Fig.~\ref{fig:merging_val} (top), where we plot those contributions
and compare their relative impact on the inclusive spectrum. Secondly, although there is an 
enhancement at high $p_T^{j_\mathrm{lead}}$ events when employing multi-jet
merging, there is also a suppression in the intermediate $p_T^{j_\mathrm{lead}}$ regime. 
In Fig.~\ref{fig:merging_val} (bottom), we observe that this suppression factor reaches  almost 
$0.5$ at $p_T^{j_\mathrm{lead}}\approx\SI[scientific-notation=false]{50}{\giga\electronvolt}$. 
The corresponding phase space does, to some extend, overlap with the high-$E_T^\mathrm{miss}$
and therefore compensates an enhancement in this region that would be
due to large $p_T^{j_\mathrm{lead}}$ configurations.

In order to demonstrate that the suppression of the intermediate
$p_T^{j_\mathrm{lead}}$ region is a genuine effect of the higher
multiplicity processes, that we include through merging, we compare our
results for the leading jet transverse momentum spectrum to a
fixed-order calculation in the bottom panel of
Fig.~\ref{fig:merging_val}. We observe a very good agreement of the
fixed order results with the multi-jet merged prediction in the
relevant regions of phase space recovering the suppression around
$p_T^{j_\mathrm{lead}}\approx\SI[scientific-notation=false]{50}{\giga\electronvolt}$.

\end{appendix}


\end{document}